# On Performance and Limitations of Quantum Wave Packet Dynamics Simulated using NISQ Hardware

Tamila Kuanysheva,[1] Jonathan Andrade-Plascencia,[2] Jayakrushna Sahoo,[1] Brian Kendrick,[3] and Dmitri Babikov[1*]

**Affiliations:**

[1]*Chemistry Department, Marquette University, Milwaukee, Wisconsin 53201-1881, USA*
[2]*Cornell University, Ithaca, New York, 14853, USA*
[3]*Los Alamos National Laboratory, Theory Division, T-1, Los Alamos, New Mexico 87545, USA*
* E-mail: dmitri.babikov@mu.edu

**Abstract:** We investigate one-dimensional quantum wave-packet dynamics using a grid-based encoding of the wave function onto qubit registers. Time evolution is implemented using a split-operator approach, with the kinetic energy operator realized via the Quantum Fourier Transform (QFT) and arbitrary discretized potentials represented by commuting Pauli-Z gates. The algorithm is benchmarked against classical grid-based simulations and validated using a noiseless Qiskit Statevector simulator before being implemented on current quantum hardware (IBM Quantum and IonQ) for qubit counts ranging from two to five. Good agreement with benchmark calculations is obtained for two- and three-qubit implementations, while deviations increase at larger qubit counts due to hardware noise and increased circuit depth. Among the devices examined, the algorithm maintains closer agreement with the benchmark on the trapped-ion IonQ Forte processor as the number of qubits increases.

## I. INTRODUCTION

The simulation of quantum dynamical processes is central to understanding a wide range of phenomena in chemistry and physics,[1–6] including molecular vibrations,[7–11] scattering events,[12–16] quantum tunneling,[17,18] nonadiabatic transitions,[19,20] and energy transfer between molecular degrees of freedom.[21] These processes are governed by the time-dependent Schrödinger equation, which describes the evolution of quantum wave packets in time. Despite its fundamental importance, accurate numerical simulation of quantum dynamics remains computationally challenging. Classical approaches based on direct propagation of the wave function suffer from the *curse of dimensionality*, where the number of grid points or basis functions required to represent the system grows exponentially with the number of degrees of freedom.[22] As a result, both grid-based and basis-set methods become rapidly intractable for systems beyond a few dimensions, particularly when long-time wave packet propagation is required.

Quantum computers provide a natural platform for simulating quantum mechanical systems because they operate according to the same underlying physical principles.[23] In particular, the state of an $n$-qubit quantum register is described by a vector in a Hilbert space of dimension $2^n$, allowing quantum systems with exponentially large state spaces to be represented using a number of qubits that grows only linearly with system size. This property makes quantum computers particularly attractive for problems affected by the curse of dimensionality. This property has motivated substantial progress in quantum algorithms for chemical physics, particularly in electronic structure problems.[24–41] Quantum simulations of real-time dynamics have also been investigated for several decades, including Fourier-transform-based approaches for wave-packet propagation and their applications to chemical dynamics.[8,10,42–47] More recently, advances in Hamiltonian simulation, including qubitization and quantum signal processing, have established asymptotically optimal algorithms for time evolution on fault-tolerant quantum computers.[48–52]

Despite these developments, relatively little is known about how accurately current noisy intermediate-scale quantum (NISQ) hardware can reproduce benchmark quantum-dynamics calculations as circuit complexity increases. While several proof-of-principle demonstrations of quantum dynamics have been reported,[8,10,42–44] systematic assessments of algorithm performance on current quantum hardware across increasing qubit counts and increasingly complex dynamical problems remain limited. This is precisely the objective of the present work.

In our previous work,[43] we simulated quantum dynamics on a quantum computer using a split-operator approach in which the kinetic energy operator was implemented via the Quantum Fourier Transform (QFT) and the potential energy operator was implemented as a quantum circuit derived from an analytic potential energy function. Although this approach successfully reproduced the dynamics of simple models on a noiseless Statevector simulator (Qiskit SDK primitives)[53], it was limited to systems with known analytic potentials and was found to be noisy for a five-qubit hardware simulation. [43] Building on this framework, we extend the methodology by replacing the analytic representation of the potential energy operator with a general Pauli-string formulation,[40,41,54] which can accommodate arbitrary discretized potentials. Using this formulation, we simulate one-dimensional wave-packet dynamics on quantum computers with grid-based encodings ranging from two to five qubits to assess how simulation accuracy changes with increasing register size. The algorithm is first validated against noiseless Statevector and subsequently executed on current NISQ hardware,[55–57] including IBM superconducting processors (Torino, Boston, and Miami) and the IonQ Forte trapped-ion device, to evaluate its performance under realistic experimental conditions. The methodology is applied to three representative problems in quantum molecular dynamics: propagation of wave packet on a flat potential, tunneling through a barrier, and oscillations in a confining potential. These scenarios correspond to free particle motion that precludes any collision process, activation barrier crossing in chemical reactions, and vibrational motion of molecular bonds, respectively.

Although QFT-based algorithms and Pauli-operator representations have been discussed extensively in the literature, experimental implementations of this framework on current quantum hardware remain relatively limited. Previous demonstrations have generally been restricted to smallest quantum registers, up to two qubits,[41,42,44,58] and, more recently, to three qubits.[47] To the best of our knowledge, applications of this framework to arbitrary discretized potentials on current quantum hardware, together with a systematic comparison against benchmark calculations as a function of qubit number, have not previously been reported.

## II. THEORY

We consider the time evolution of a one-dimensional wave function $\psi(r,t)$, where $r$ represents a spatial coordinate such as a molecular bond length or reaction coordinate. The continuous coordinate is discretized on a uniform grid of $M = 2^n$ points spanning the interval $[r_{min}, r_{max}]$, with spacing $\Delta r = (r_{\max} - r_{\min})/M$. This corresponds to a discrete-variable representation (DVR)[12] of a wave function. This formulation is closely related to pseudospectral methods, where a continuous wave function is represented on a spatial grid and differential operators are evaluated efficiently using spectral (Fourier-based) representations.[59,60] The wave function is represented in the discrete position basis as $\psi_m(t) = \psi(r_m, t)$, where $r_m$ denotes the grid points. The discretized wave function is encoded on an $n$-qubit register using a direct mapping between grid points and computational basis states, $|m\rangle$, such that

$$|\psi(t)\rangle = \sum_{m=0}^{M-1} \psi_m(t)\,|m\rangle. \qquad (1)$$

The integer index $m$ corresponds to the binary representation of the qubit register. In tensor-product form, the basis state is written as

$$|m\rangle = |q_{n-1}\rangle \otimes |q_{n-2}\rangle \otimes \cdots \otimes |q_0\rangle, \qquad (2)$$

where $q_j \in \{0,1\}$ denotes the state of qubit $j$. Throughout this work we follow the little-endian convention used in Qiskit, where qubit 0 represents the least significant bit in the binary encoding of $m$.[61] For example, a $n = 2$ qubits provides $M = 4$ basis states that correspond to four grid points:

$$|00\rangle \to r_0, |01\rangle \to r_1, |10\rangle \to r_2, |11\rangle \to r_3.$$

In this work, simulations are performed for quantum registers containing between two and five qubits, corresponding to spatial grids of $M = 4$ to $M = 32$ points. To facilitate comparison across different qubit numbers, all results are mapped to a common 32-point grid.

Throughout this work, atomic units are used ($\hbar = 1$). The time evolution of the wave function is governed by the time-dependent Schrödinger equation. Formally, the wave function at time $t$ is obtained by applying the time-evolution operator to the initial state,

$$|\psi(t)\rangle = e^{-iHt}|\psi(0)\rangle. \qquad (3)$$

A general Hamiltonian can be encoded on a quantum computer using a decomposition into sequence of tensor products of Pauli gates; however, this representation can become challenging in practice, as the number of terms in such a decomposition grows exponentially with the number of qubits.[40] The Hamiltonian operator can be written as $\hat{H} = \hat{T} + \hat{V}$, where $\hat{T}$ and $\hat{V}$ are the kinetic and potential energy operators, respectively. In position representation, the potential is given by $\hat{V} = V(r)$, while the kinetic operator $\hat{T} = \hat{p}^2/2\mu$, where $\hat{p}$ is the momentum operator and $\mu$ is the reduced mass.

To simulate time evolution numerically, we employ the Split-Operator method.[45,62–64] The total propagation time $t_{\text{fin}}$ is divided into $N$ small time steps of duration $\Delta t$, $t_{\text{fin}} = N\Delta t$. The global propagator can then be written as a product of short-time propagators, $e^{-i\hat{H}t_{\text{fin}}} = \left(e^{-i\hat{H}\Delta t}\right)^N$. For sufficiently small $\Delta t$, each short-time propagator can be approximated using a first-order Trotter decomposition,[65]

$$e^{-i\hat{H}\Delta t} \approx e^{-i\hat{V}\Delta t}\, e^{-i(\hat{p}^2/2\mu)\Delta t} + O(\Delta t^2) \qquad (4)$$

which separates the potential and kinetic contributions to the time evolution.

In the grid-based encoding described above, an $n$-qubit register corresponds to $M = 2^n$ spatial grid points. The wave function is therefore represented by a state vector of dimension $2^n$,

$$|\psi\rangle = \begin{pmatrix} \psi_0 \\ \psi_1 \\ \vdots \\ \psi_{M-1} \end{pmatrix}.$$

Consequently, quantum operators acting on this state, including the Hamiltonian and its potential component, are represented by matrices of dimension $2^n \times 2^n$. The potential energy operator is diagonal in the position basis and can be written as

$$\hat{V} = \sum_{m=0}^{M-1} V_m \, |m\rangle\langle m|, \tag{5}$$

where $V_m = V(r_m)$ is the value of potential at grid point $r_m$. The matrix representation of this operator therefore contains the potential values along the diagonal and zeros elsewhere.

In our previous work, the potential energy part of the time propagation operator $e^{-i\hat{V}\Delta t}$ was implemented as quantum circuit constructed specifically for a given model system based on the analytic functional form of its potential energy $V(r)$ such as harmonic oscillator or a double-well potential.[43] While efficient for simple model potentials, this approach is limited to potentials with convenient analytic representations. Here we adopt a more general approach by expressing the potential energy operator as a linear combination of tensor products of Pauli operators $X, Y, Z$ and identity $I$.[40,41,54]

For each qubit the set of Hermitian operators $\{I, X, Y, Z\}$ represents a complete operator basis:

$$I = \begin{pmatrix} 1 & 0 \\ 0 & 1 \end{pmatrix}, X = \begin{pmatrix} 0 & 1 \\ 1 & 0 \end{pmatrix}, Y = \begin{pmatrix} 0 & -i \\ i & 0 \end{pmatrix}, Z = \begin{pmatrix} 1 & 0 \\ 0 & -1 \end{pmatrix}.$$

For $n$ qubits, a complete operator basis is formed by all possible tensor products of these gates for all qubits, with each element of this basis referred to as a *Pauli string*:

$$P = P_{n-1} \otimes P_{n-2} \otimes \cdots \otimes P_0, \tag{6}$$

where $P_j \in \{I, X, Y, Z\}$ acts on qubit $j$. Because each qubit contributes four possible operators, the total number of Pauli strings acting on $n$ qubits is $4^n$. Any operator acting on the $2^n$-dimensional Hilbert space can therefore be expanded as $\hat{V} = \sum_{k=1}^{4^n} c_k P_k$, where $P_k$ enumerates the Pauli strings and $c_k$ are real coefficients. The coefficients are obtained using the orthonormality condition of Pauli operators, $c_k = \mathrm{Tr}(P_k V)/2^n$ where $V$ is the matrix representation of $\hat{V}$. A detailed derivation for one- and two-qubit implementations is provided in our previous work,[44] and the method generalizes straightforwardly to more qubits.

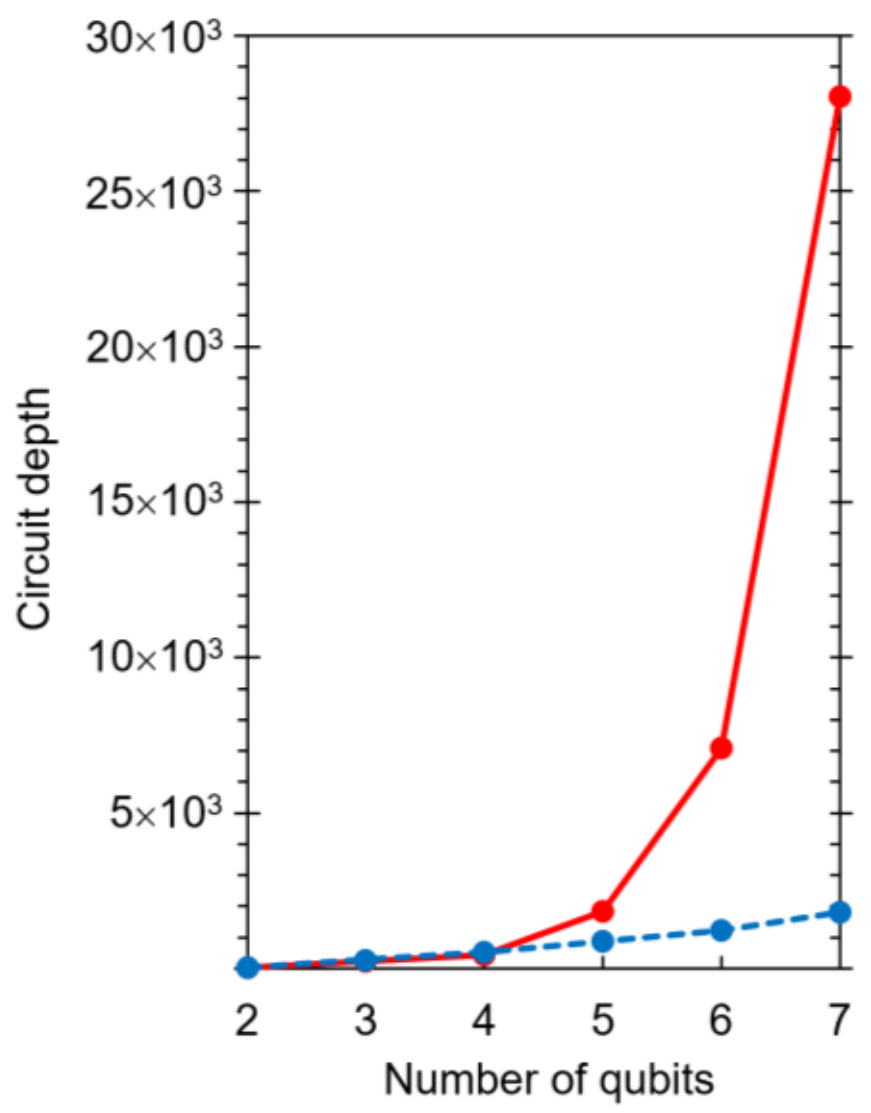

**Figure 1.** Circuit depth as a function of the number of qubits (2–7) for a single time step of the time-evolution operator with a harmonic oscillator potential. Red curve corresponds to a full Pauli decomposition of the Hamiltonian, while blue curve represents the approach used in this work, where only the potential energy operator is expanded in Pauli-Z strings and the kinetic energy operator is implemented via the Quantum Fourier Transform (QFT). Circuits were transpiled for the IBM Torino backend.

Decomposition of a general Hamiltonian matrix requires up to $4^n$ terms, leading to exponential growth with increasing number of qubits. In the present case, however, only the potential energy operator, diagonal in the position basis, is represented by Pauli operators. Importantly, its expansion contains only tensor products of $I$ and $Z$ operators and only $2^n$ terms in total.[46] Each Pauli string acts on all $n$ qubits, with some qubits carrying $Z$ operators and others the identity $I$. The potential operator can therefore be written as a sum of commuting Pauli-Z strings,

$$\hat{V} = \sum_k c_k \; Z_{n-1}^{(k)} \otimes Z_{n-2}^{(k)} \otimes \cdots \otimes Z_0^{(k)} \tag{7}$$

Overall, the potential part of the time evolution operator is implemented as

$$e^{-i\hat{V}\Delta t} = \prod_k e^{-ic_k P_k \Delta t}. \tag{8}$$

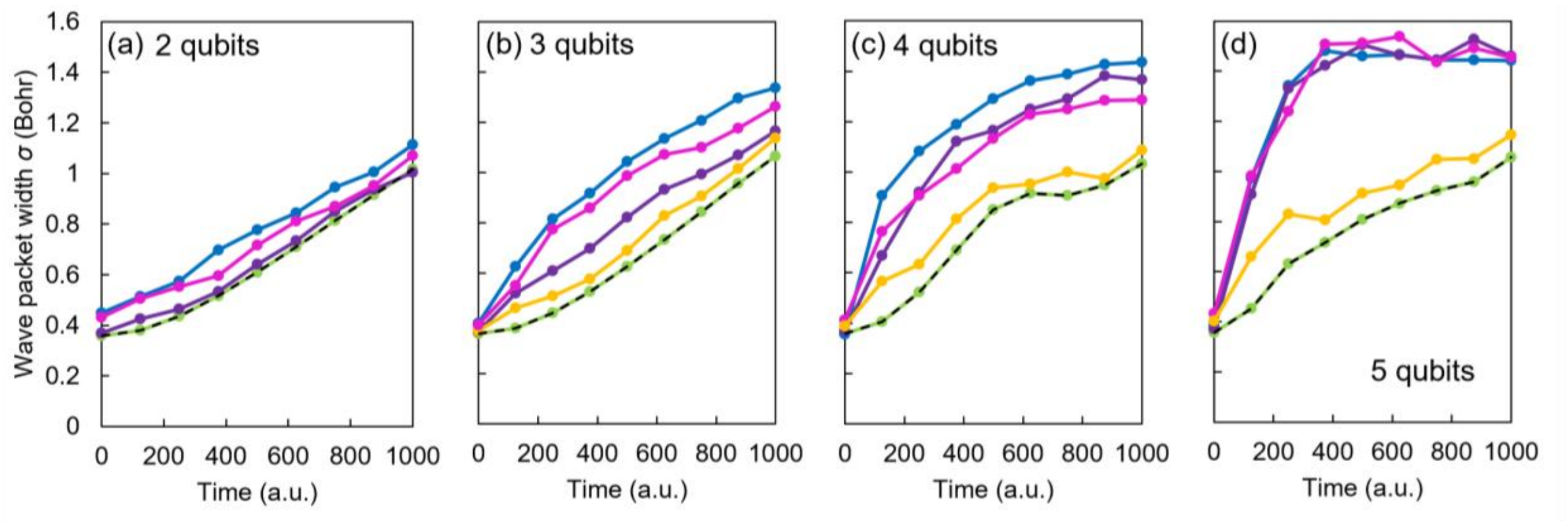

**Figure 2.** Time evolution of free-particle wave packet. Panels (a)-(d) correspond to simulations with 2 to 5 qubits, respectively. Symbols correspond to eight individual time-steps. The dashed black curves represent the classical benchmark, while green curves show wave-packet widths obtained using the noiseless Qiskit Statevector simulator. Blue, purple, and pink curves correspond to results obtained on IBM quantum processors: IBM Torino (Heron architecture), IBM Boston (updated Heron architecture), and IBM Miami (Nighthawk architecture), respectively. Yellow curves in panels (b-d) show results obtained using IonQ Forte 1 trapped-ion quantum computer.

Another recent work[6,58] took advantage of the operator structure in quantum dynamics simulations through singular value decomposition (SVD), where a linear operator is factorized into unitary transformations and a diagonal operator that can be efficiently implemented in a dilated space. Although the present approach does not employ SVD, both methods benefit from structured representations of operators that reduce circuit complexity.

The kinetic energy operator is implemented using the Quantum Fourier Transform (QFT), which maps the wave function to the momentum basis where $\hat{T} = \hat{p}^2/2\mu$ is diagonal. The QFT is implemented as a quantum circuit composed of elementary gates, including Hadamard gates, controlled-phase gates, and qubit swaps.[43,66] The propagation is given by [45,65,67,68]

$$e^{-i\hat{T}\Delta t} = \mathrm{QFT}^{-1}\, e^{-i(p^2/2\mu)\Delta t}\, \mathrm{QFT}. \qquad (9)$$

The QFT circuit requires only $O(n^2)$ elementary gates for an $n$-qubit register. This compares favorably with the original discrete Fourier transform, $O(2^{2n})$, and fast Fourier transform, $O(n2^n)$, algorithms.[66] This approach is closely related to Fourier pseudospectral methods, where transformations between coordinate and spectral representations are used to evaluate each operator in the basis where it is most naturally represented.[59] By transforming the wave function to the momentum representation, where the kinetic operator is diagonal, the evolution operator can be applied with circuit depth that scales more efficiently with register size. compares circuit depth as a function of the number of qubits for a single time step of the time-evolution operator with a harmonic oscillator potential, using a full Pauli decomposition of the Hamiltonian and the present approach, in which only the potential operator is expanded in Pauli strings while the kinetic operator is implemented via the QFT. In a five-qubit case, the QFT-based approach reduces the circuit depth from 1845 to 889, while for a seven-qubit case, the depth decreases from 28057 to 1797, demonstrating the substantial reduction in circuit complexity. The quantum circuit for a representative four-qubit Pauli-$Z$ string contributing to the potential energy operator is provided in Figure S1 of the supporting material. The quantum circuit for the QFT is shown in Figure S2.

## III. RESULTS AND DISCUSSION

We perform simulations of three representative one-dimensional quantum dynamics problems - free particle propagation, tunneling through a barrier, and harmonic oscillator - using registers of 2 to 5 qubits, corresponding to spatial grids of $M = 4$ to 32 points. In all cases, the initial state is prepared as a step-like wave packet localized within the second quarter of the grid, with uniform amplitude over $M/4 \leq m < M/2$. More details about wave packet initialization are given in the supplementary material. Time evolution is carried out for eight time-steps.

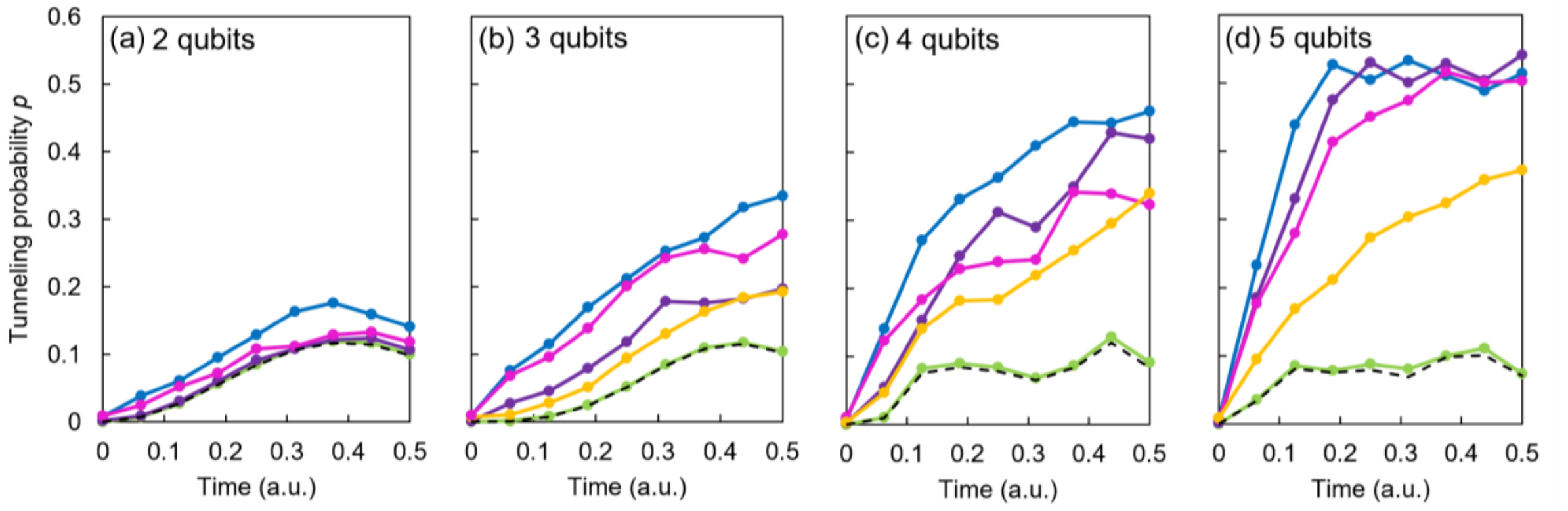

**Figure 3.** Quantum tunneling through a rectangular barrier. Panels (a)-(d) correspond to simulations with 2 to 5 qubits, respectively. Symbols correspond to eight individual time-steps. The dashed black curves represent the classical benchmark, while green curves show wave-packet widths obtained using the noiseless Qiskit Statevector simulator. Blue, purple, and pink curves correspond to results obtained on IBM quantum processors: IBM Torino (Heron architecture), IBM Boston (updated Heron architecture), and IBM Miami (Nighthawk architecture), respectively. Yellow curves in panels (b-d) show results obtained using IonQ Forte 1 trapped-ion quantum computer.

We first consider free particle propagation on a flat potential, where the Hamiltonian contains only the kinetic energy operator. The spatial domain is defined over $[0, 5]$ Bohr, with time step $\Delta t = 125$ a.u. and total evolution time $t_{\text{fin}} = 1000$ a.u. The reduced mass of the OH radical ($\mu = 0.9412$ amu) is used, consistent with prior quantum computing studies of this system.[69] During propagation, the wave packet broadens as expected, with the width $\sigma = \sqrt{\langle r^2 \rangle - \langle r \rangle^2}$ increasing from 0.36 Bohr at $t = 0$ to 1.07 Bohr at the final time, with minor variations due to grid resolution.

The results of these simulations are presented in Figure 2 of the main text and Table S1 of the supplementary material. The dashed black curve represents the classical benchmark, while the green curve shows the results obtained using the noiseless Qiskit Statevector simulator.[53] Unless otherwise noted, all hardware results were obtained using 4096 measurement shots for the IBM quantum processors and 1024 measurement shots for the IonQ Forte 1 trapped-ion quantum processor. For the two-qubit case (Figure 2a), all IBM quantum hardware platforms reproduce the qualitative broadening behavior, with varying levels of accuracy. The IBM Torino backend (blue curve, now retired) incorporates Heron r1 processor and captures the overall trend but exhibits a systematic offset from the benchmark. The newer IBM Miami (pink curve), that uses Nighthawk r1 processor and an architecture with enhanced connectivity, shows modest improvement, while the IBM Boston backend (purple curve) with newer Heron r3 processor shows near-quantitative agreement with the emulator, indicating high-fidelity performance in this two-qubit test run.

For three qubits (Figure 2b), a similar trend is observed, with IBM Boston showing better agreement with the benchmark than IBM Miami and IBM Torino. However, deviations from the benchmark become more pronounced for all superconducting devices. In contrast, the trapped-ion IonQ Forte 1 processor (yellow curve) shows significantly improved agreement with the benchmark, consistent with the higher fidelity of two-qubit gates and all-to-all connectivity available in trapped-ion quantum architectures.[70–72]

As the number of qubits in a test increases further (panels c and d), the accuracy of IBM quantum hardware platforms deteriorates. For four qubits (Figure 2c), IBM Boston and Miami show better agreement with the benchmark than Torino, although all devices exhibit noticeable deviations from the benchmark. In contrast, IonQ Forte maintains close agreement with the emulator throughout the time evolution. At five qubits (Figure 2d), all IBM backends exhibit a rapid loss of fidelity: by the third time step, the wave function collapses to an approximately uniform distribution over all basis states and remains in this state for the remainder of the propagation, indicating a loss of visible wave-packet structure. IonQ Forte again preserves good agreement

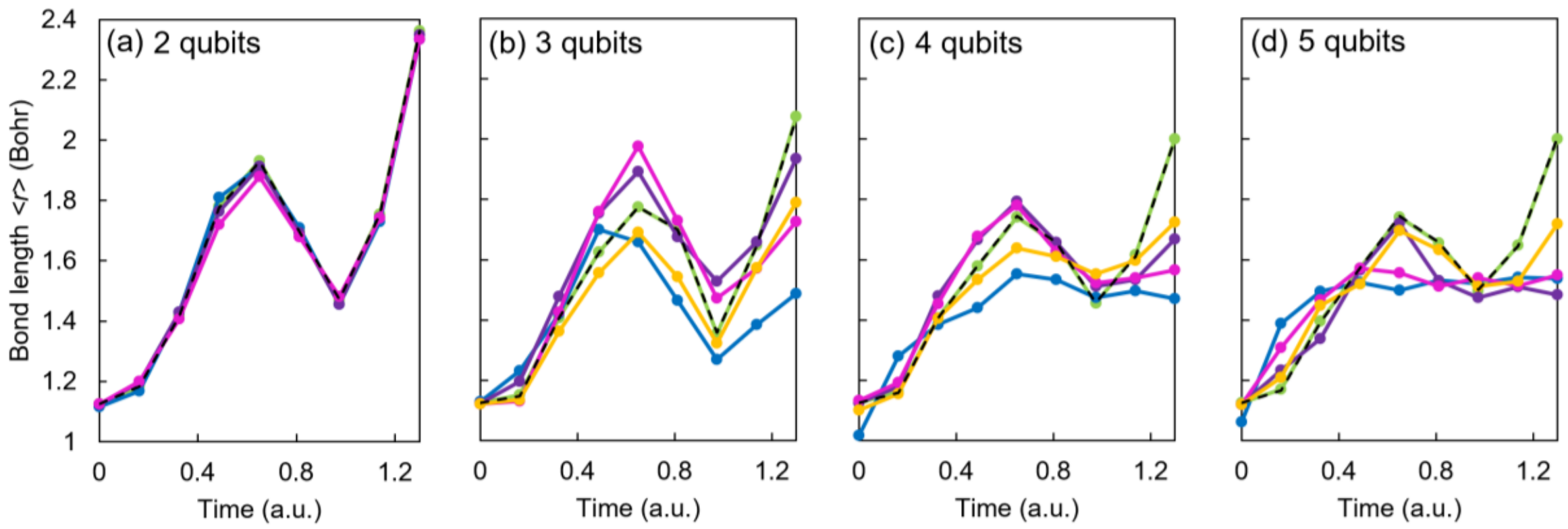

**Figure 4.** Time evolution of a wave packet in a harmonic oscillator potential. Panels (a)-(d) correspond to simulations with 2 to 5 qubits, respectively. Symbols correspond to eight individual time-steps. The dashed black curves represent the classical benchmark, while green curves show wave-packet widths obtained using the noiseless Qiskit Statevector simulator. Blue, purple, and pink curves correspond to results obtained on IBM quantum processors: IBM Torino (Heron architecture), IBM Boston (updated Heron architecture), and IBM Miami (Nighthawk architecture), respectively. Yellow curves in panels (b-d) show results obtained using IonQ Forte 1 trapped-ion quantum computer.

with the benchmark, maintaining the wave-packet features over the full evolution.

We next consider quantum tunneling through a barrier problem, in which the Hamiltonian includes both kinetic and potential energy contributions. The spatial grid spans the interval $[0, 4]$ Bohr, and a piecewise-constant double-well potential is defined with minimum value $V_{\min} = -6$ Hartree,

$$V(r_m) = \begin{cases} 0, & 0 \leq m < M/4 \\ V_{\min}, & M/4 \leq m < M/2 \\ 0, & M/2 \leq m < 3M/4 \\ V_{\min}, & 3M/4 \leq m < M. \end{cases}$$

The initial wave packet is uniformly distributed within the left well. Time evolution is performed using a time step $\Delta t = 0.0625$ a.u. up to a final time $t_{\text{fin}} = 0.5$ a.u., with the reduced mass set to $\mu = 1$ a.m.u. The tunneling probability is quantified as the loss of population from the initial well,

$$p = 1 - \sum_{m=M/8}^{5M/8-1} |\psi(r_m)|^2.$$

The results are presented in Figure 3 and Table S2 of the supplementary material. The probability of finding the particle in the right well increases monotonically during the evolution, reaching approximately 7–10% at the final time. The overall hardware-dependent behavior follows the same trend observed for free-particle propagation (Figure 2), with deviations from the benchmark increasing with the number of qubits and improved performance observed for newer superconducting devices and trapped-ion hardware. Among the tested platforms, IonQ Forte consistently shows closer agreement with the benchmark than the IBM devices, with the difference being most pronounced in the five-qubit case.

We next consider a harmonic oscillator potential, which introduces a smooth quadratic potential energy landscape in contrast to the piecewise-constant barrier considered previously. The spatial grid spans the interval $[0, 3]$ Bohr, and a parabolic potential centered at $r_{\text{eq}} = 1.5$ Bohr with frequency $\omega = 1$ Hartree is employed,

$$V(r_m) = \frac{1}{2}\mu\omega^2(r_m - r_{\text{eq}})^2,$$

where $\mu = 1$ a.m.u. Time evolution is performed with a time step $\Delta t = 0.1625$ a.u. up to a final time $t_{\text{fin}} = 1.3$ a.u. The expectation value of the position is evaluated as

$$\langle r \rangle = \sum_{m=0}^{M-1} r_m |\psi(r_m)|^2.$$

The results are shown in Figure 4 and Table S3 of the supplementary material. The expectation value $\langle r \rangle$ exhibits oscillatory behavior, reflecting vibrational motion of the wave packet in the harmonic potential. The wave packet initially moves away from the equilibrium position, followed by a reversal of motion, consistent with the restoring force of the harmonic oscillator.

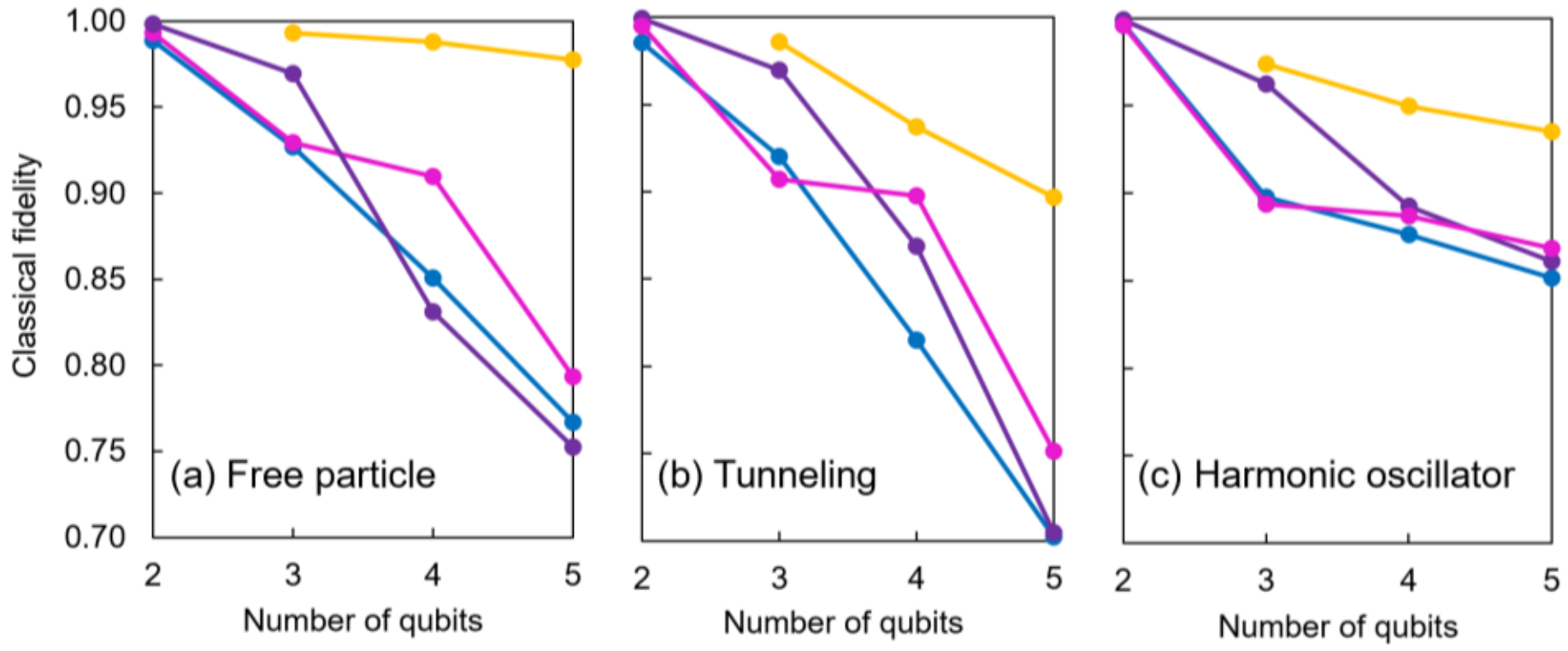

**Figure 5.** Average classical Bhattacharyya fidelity between probability distributions obtained on quantum hardware and the corresponding noise-free benchmark simulations as a function of the number of qubits (2-5). Panels (a)–(c) correspond to free-particle wave-packet propagation, quantum tunneling through a rectangular barrier, and wave-packet evolution in a harmonic oscillator potential, respectively. Blue, purple, pink, and yellow curves represent results obtained on the IBM Torino (Heron architecture), IBM Boston (updated Heron architecture), IBM Miami (Nighthawk architecture), and IonQ Forte 1 trapped-ion quantum processor, respectively.

For the two-qubit implementation (Figure 4a), all IBM superconducting backends show close agreement with the classical benchmark. For the three-qubit case (Figure 4b), IBM Boston, IBM Miami, and IonQ Forte show closer agreement with the benchmark compared to IBM Torino. As the number of qubits increases further, deviations from the benchmark become progressively more pronounced. For four qubits (Figure 4c), IBM Torino exhibits rapid degradation: by the fourth time step, the wave function collapses toward an approximately uniform distribution over all basis states and remains in this state for the remainder of the evolution. IBM Miami, IBM Boston, and IonQ Forte remain in closer agreement with the classical benchmark for a longer duration, although deviations begin to emerge toward the end of the propagation. In the five-qubit case (Figure 4d), all superconducting backends exhibit substantial deviations from the benchmark. IBM Torino and IBM Miami deviate as early as the second time step, while IBM Boston maintains reasonable agreement for approximately half of the evolution before diverging at later times. The trapped-ion IonQ Forte processor remains in close agreement with the benchmark throughout most of the propagation, with noticeable deviations appearing only near the end of the simulation.

To provide an additional quantitative measure of agreement between the hardware results and the benchmark calculations, we computed the classical Bhattacharyya fidelity[73,74] between the probability distributions obtained on quantum hardware and those produced by the noiseless Qiskit Statevector[53] simulator at each propagation time step,

$$F(t) = \left(\sum_{i=1}^{2^n} \sqrt{p_i(t) q_i(t)}\right)^2. \qquad (10)$$

The fidelity values were then averaged over the entire time evolution according to

$$\bar{F} = \frac{1}{N} \sum_{t=0}^{N-1} F(t), \qquad (11)$$

where $N$ is the number of propagation time steps, and $p_i(t)$ and $q_i(t)$ are the probabilities of basis state $i$ at time $t$ obtained from the hardware and benchmark simulations, respectively. The value of $\bar{F} = 1$ corresponds to perfect agreement with the benchmark. The resulting average fidelities are presented in Figure 5 of the main text and Table S4 of the supplementary material. The fidelity values at each propagation time step for all benchmark problems are provided in Tables S5-S7 of the supplementary material. In general, the fidelity decreases as the propagation proceeds, consistent with the accumulation of hardware noise and errors associated with increasing circuit depth during the time evolution.

For all three benchmark problems, the fidelity decreases systematically with increasing qubit number, reflecting the accumulation of hardware errors as circuit size and depth increase. For the free-particle problem (Figure 5a), all IBM hardware platforms exhibit excellent agreement with the benchmark for two-qubit simulations, with average fidelities above 0.98. At three qubits, the fidelities remain above 0.92 for all devices. As the "spatial-grid resolution increases further, the IBM superconducting processors show a gradual reduction in fidelity, reaching values between 0.75 and 0.79 for five qubits. In contrast, IonQ Forte maintains fidelities above 0.97 throughout the entire range of qubit numbers considered.

A similar trend is observed for the tunneling problem (Figure 5b). Two-qubit simulations again exhibit nearly perfect agreement with the benchmark, while three-qubit fidelities remain close to or above 0.90 for all platforms. For four- and five-qubit calculations, the fidelity decreases more rapidly than in the free-particle case, particularly for the superconducting processors. IonQ Forte again maintains the highest agreement with the benchmark, with an average fidelity of 0.897 for the five-qubit simulation.

For the harmonic oscillator (Figure 5c), two-qubit fidelities exceed 0.99 on all platforms. Although the fidelity decreases with increasing qubit number, the reduction is less pronounced than in the tunneling case. At five qubits, the IBM processors yield fidelities between 0.85 and 0.87, while IonQ Forte retains an average fidelity of 0.94. Overall, these quantitative fidelity measurements are consistent with the behavior observed in Figures 2-4. They demonstrate that all hardware platforms accurately reproduce the benchmark dynamics for small quantum registers, while increasing circuit depth leads to progressively larger deviations from the benchmark. Among the devices examined, the trapped-ion IonQ Forte processor shows the smallest decrease in fidelity with increasing qubit number.

In this work, we evaluate the performance of a split-operator framework for quantum wave-packet dynamics on current NISQ hardware. The kinetic energy operator is implemented via the QFT, while the potential-energy operator is represented using commuting Pauli-Z strings, providing a framework that can accommodate arbitrary discretized potentials. arbitrary discretized potentials are represented using commuting Pauli-Z strings. Here, we demonstrate the approach for three representative problems with model analytically defined potentials: free wave-packet propagation, tunneling through a barrier, and vibrational motion in a harmonic potential. Future work will extend this approach to more complex, arbitrary discretized potential-energy landscapes.

The simulations use one-dimensional spatial grids encoded with two to five qubits. The quantum-circuit results were benchmarked against classical grid-based calculations and a noiseless Qiskit Statevector simulation before execution on IBM superconducting processors (Torino, Boston, and Miami) and the IonQ Forte trapped-ion device. Good agreement with classical calculations is observed for two- and three-qubit simulations, as confirmed by the classical Bhattacharyya fidelity analysis, while larger qubit counts reveal the increasing impact of hardware noise and circuit-depth limitations on algorithm performance. Among the devices examined, the trapped-ion IonQ Forte processor maintains closer agreement with benchmark calculations at larger qubit counts, consistent with its high two-qubit gate fidelities and all-to-all qubit connectivity.[70–72] Approximate QFT implementations, which omit the smallest controlled-phase rotations, could be explored in future work as a strategy for reducing circuit depth and potentially improving performance on noisy quantum hardware. More broadly, potential errors associated with QFT implementation motivate exploring alternative circuit-synthesis strategies, such as Quantum Shannon Decomposition (QSD), which can be used to synthesize the time-evolution operator without explicitly implementing the QFT-based split-operator sequence.[75]

Overall, the results demonstrate that despite its limitations, the split-operator approach can accurately reproduce quantum wave-packet dynamics on current quantum hardware for a small number of qubits, highlighting the practical constraints imposed by present-day NISQ devices.

## SUPPLEMENTARY MATERIAL

The supplementary material includes Figs. S1 and S2, which present the quantum circuits for the four-qubit Pauli-$Z$ string $Z \otimes Z \otimes Z \otimes Z$ and the QFT, respectively, as well as details of the initial state preparation for the time-evolution simulations. It also includes Table S4, reporting the average classical

Bhattacharyya fidelities between the quantum hardware results and the noiseless Statevector benchmark for all benchmark problems and qubit numbers, and Tables S5–S7, which provide the corresponding fidelity values at each propagation time step for the free-particle, tunneling, and harmonic oscillator simulations, respectively. The Qiskit code used to perform the five-qubit simulations is also provided within the supplementary material.

## ACKNOWLEDGMENTS

This research was supported by NSF ExpandQISE program, grant number OSI/OMA-2328489. TK acknowledges the support of Quantum Computing Summer School at LANL, Texas Quantum Winter School and IonQ Research Credits Program. BKK acknowledges that part of this work was done under the auspices of the U.S. DoE under Project No. 20240256ER of the LDRD program at LANL, operated by Triad National Security for NNSA (Contract 89233218CNA000001).

## AUTHOR DECLARATIONS

### Conflict of Interest

The authors have no conflicts to disclose.

### Author Contributions

**Tamila Kuanysheva**: Conceptualization (equal); Data curation (lead); Formal analysis (lead); Investigation (lead); Methodology (equal); Software (equal); Validation (lead); Visualization (lead); Writing – original draft (lead); Writing – review & editing (equal). **Jonathan Andrade-Plascencia**: Methodology (equal); Software (equal); Validation (supporting); Visualization (supporting); Writing – review & editing (equal). **Jayakrushna Sahoo:** Methodology (supporting); Validation (supporting); Visualization (supporting); Writing – review & editing (equal). **Brian Kendrick**: Conceptualization (equal); Funding acquisition (equal); Methodology (equal); Project administration (equal); Resources (equal); Supervision (equal); Writing – review & editing (equal). **Dmitri Babikov**: Conceptualization (equal); Funding acquisition (equal); Methodology (equal); Project administration (equal); Resources (equal); Supervision (equal); Writing – review & editing (equal).

## DATA AVAILABILITY

The data that supports the findings of this study and the Qiskit code used to generate and analyze the results are available within the supplementary material file.